\documentclass[]{spie}  

\usepackage{amsmath,amsfonts,amssymb}
\usepackage{graphicx}
\usepackage{textcomp}

\usepackage[colorlinks=true, allcolors=blue]{hyperref}

\title{Testing the X-IFU calibration requirements: an example for quantum efficiency and energy resolution}

\author[a]{Edoardo Cucchetti}
\author[a]{Fran\c{c}ois Pajot}
\author[a]{Etienne Pointecouteau}
\author[b]{Philippe Peille}
\author[a]{Gabriele Betancourt-Martinez}
\author[c,d]{Stephen J. Smith}
\author[e,f]{Marco Barbera}
\author[c]{Megan E. Eckart}
\author[c]{Simon R. Bandler}
\author[c]{Caroline A. Kilbourne}
\author[g]{Massimo Cappi}
\author[a]{Didier Barret}

\affil[a]{\normalsize IRAP, Universit\'{e} de Toulouse, CNRS, CNES, 9 Av. du Colonel Roche, 31400 Toulouse, France}
\affil[b]{CNES, 18 Avenue Edouard Belin 31400 Toulouse, France}
\affil[c]{NASA/Goddard Space Flight Center, 8800 Greenbelt Rd., Greenbelt, MD 20771, United States}
\affil[d]{CRESST and University of Maryland, Baltimore County, MD 21250, USA; University of Maryland, College Park, MD 20742,USA}
\affil[e]{Universit\`{a} degli Studi di Palermo, Dipartimento di Fisica e Chimica,  Via Archirafi 36, 90123, Palermo, Italy}
\affil[f]{Istituto Nazionale di Astrofisica, Osservatorio Astronomico di Palermo, \protect \\ Piazza del Parlamento 1, 90134 Palermo, Italy}
\affil[g]{INAF-IASF Bologna, Via Gobetti 101, 40129 Bologna, Italy}

\authorinfo{Further author information: (Send correspondence to Edoardo Cucchetti)\\  Edoardo Cucchetti: E-mail: edoardo.cucchetti@irap.omp.eu}

\begin{document} 
\maketitle

\begin{abstract}
With its array of 3840 Transition Edge Sensors (TESs) operated at 90~mK, the X-Ray Integral Field Unit (X-IFU) on board the ESA L2 mission \textit{Athena} will provide spatially resolved high-resolution spectroscopy (2.5~eV FWHM up to 7~keV) over the 0.2 to 12~keV bandpass. The in-flight performance of the X-IFU will be strongly affected by the calibration of the instrument. Uncertainties in the knowledge of the overall system, from the filter transmission to the energy scale, may introduce systematic errors in the data, which could potentially compromise science objectives -- notably those involving line characterisation e.g. turbulence velocity measurements -- if not properly accounted for.  Defining and validating calibration requirements is therefore of paramount importance.  In this paper, we put forward a simulation tool based on the most up-to-date configurations of the various subsystems (e.g. filters, detector absorbers) which allows us to estimate systematic errors related to uncertainties in the instrumental response. Notably, the effect of uncertainties in the energy resolution and of the instrumental quantum efficiency on X-IFU observations is assessed, by taking as a test case the measurements of the iron K complex in the hot gas surrounding clusters of galaxies. In-flight and ground calibration of the energy resolution and the quantum efficiency is also addressed. We demonstrate that provided an accurate calibration of the instrument, such effects should be low in both cases with respect to statistics during observations.
\end{abstract}


\section{Introduction}

The X-ray Integral Field Unit (X-IFU) [\citen{Barret2016XIFU, Pajot2018XIFU}] on board the future European X-ray observatory \textit{Athena} [\citen{Nandra2013Athena}] will provide a new look at the X-ray sky and study the hot and energetic Universe with unprecedented capabilities.  With a spectral resolution of 2.5~eV required up to 7~keV and a 5'' spatial resolution over a 5' equivalent diameter field-of-view, the X-IFU will provide spatially-resolved high-resolution spectroscopy in the soft X-ray band (0.2~--~12~keV). This will be achieved using an array of $\sim$ 3840  superconducting Transition Edge Sensors (TESs) [\citen{Smith2016Pix}] micro-calorimeters operated at 90~mK, which will allow individual energy measurements for each pixel.

The calibration of the instrument will be critical to achieve its challenging science objectives. Notably, the instrumental response needs to be known with a high level of accuracy to understand the future observations and to maximise the performances of the instrument. This response will be calibrated on the ground and modelled by specific response files for each pixel, which will contain information on the effective area and energy redistribution of the detector. However, an accurate per pixel calibration may not always be possible for every physical parameter (e.g., per-pixel measurements of the stopping power) and this may cause slight systematic effects in the data when pixels are considered individually. Numerical simulations are therefore required to verify the calibration accuracy required on the instrumental parameters and to quantify the corresponding impact on the driving science objectives of the X-IFU.

In this paper, we present a numerical approach to verify calibration requirements for the X-IFU. Particular attention is given to the calibration of the instrumental response. After explaining the motivation for this approach and the calibration strategy for the detector response (Sect.~\ref{sec:resp}), we present the related simulation pipeline used to conduct theses studies (Sect.~\ref{sec:num}). This tool is applied to two examples to assess the systematic errors related to uncertainties in the quantum efficiency (Sect.~\ref{sec:qe}) and of the energy resolution (Sect.~\ref{sec:res}). In both cases, the realistic variations expected in instrumental parameters are evaluated in regard to the driving science case of the iron K complex measurement in the core of galaxy clusters. 

\section{X-IFU instrumental response calibration}
\label{sec:resp}

Whenever an astrophysical source is observed with the X-IFU, its spectrum will be convolved with the response of the instrument. This transfer function will depend on a large number of parameters. For X-ray instruments, the response is modelled using specific response files, divided into two parts for each detector:
\begin{figure}
\centering
\includegraphics[width=0.49\textwidth]{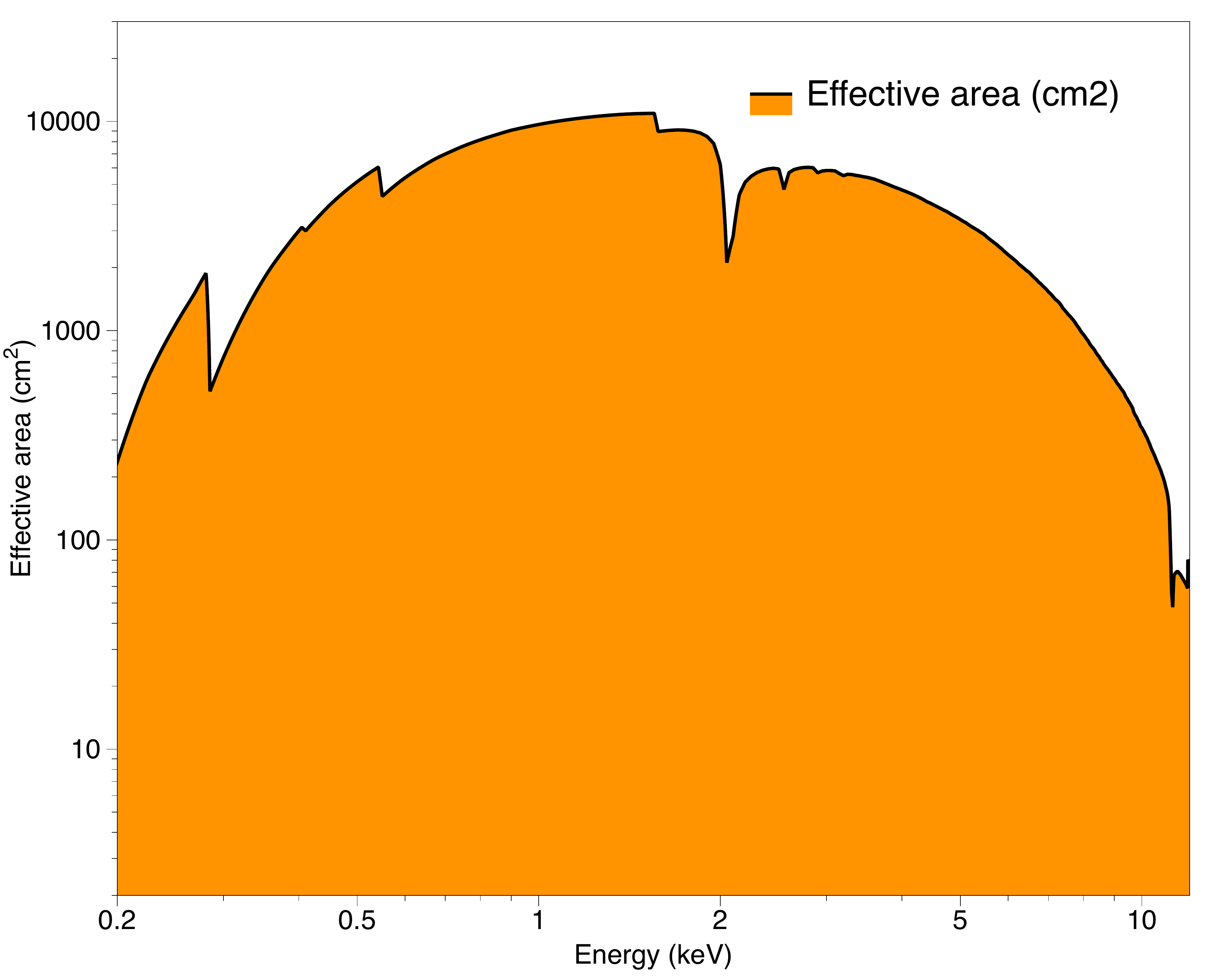}
\includegraphics[width=0.49\textwidth]{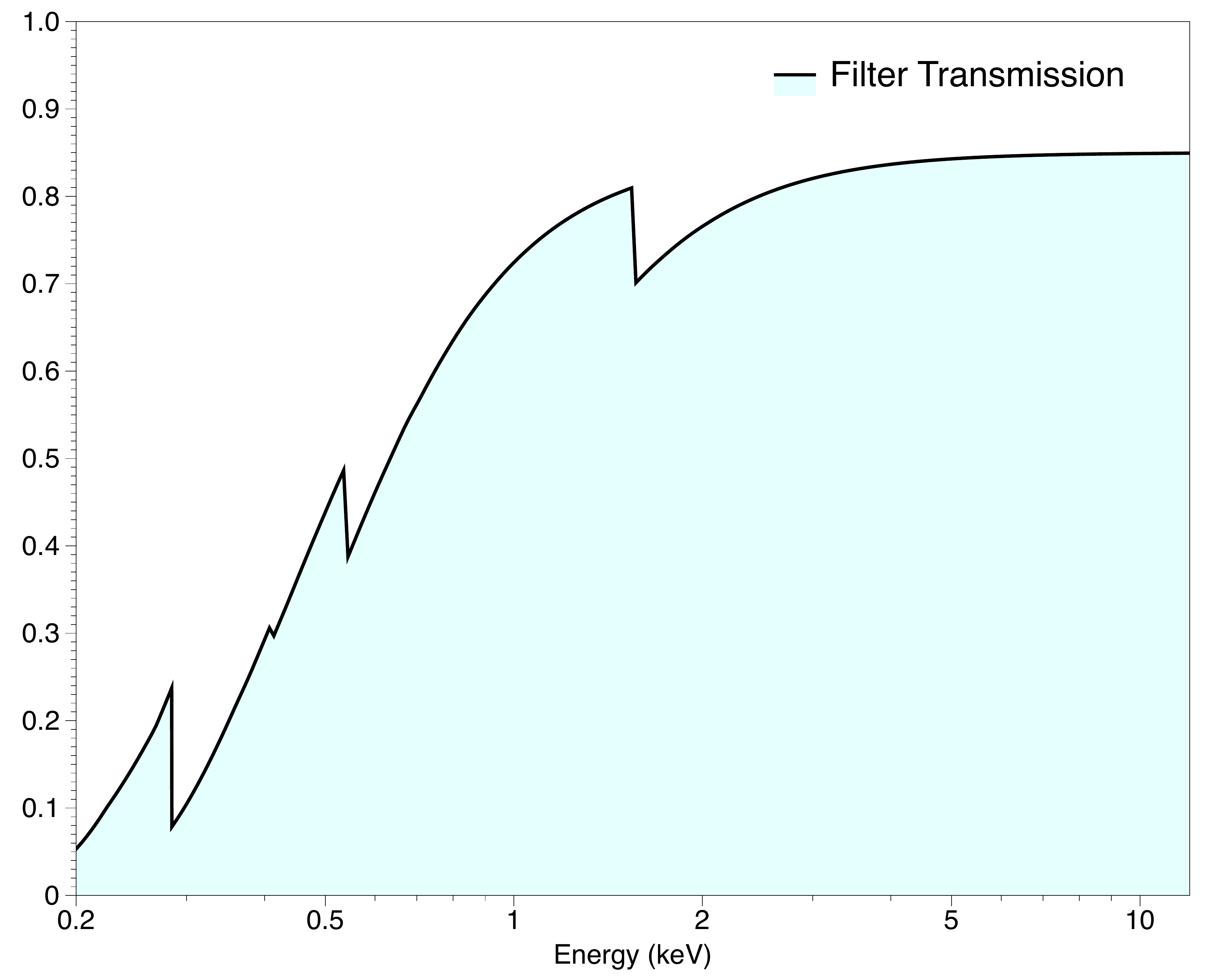}
\caption{(\textit{Left}) Total effective area for the 15-row mirrow configuration of the X-IFU (cm$^{2}$) as a function of energy (keV). (\textit{Right}) Total transmission of the thermal filters placed on the aperture cylinder of the instrument (courtesy M. Barbera) [\citen{Barbera2016Filters}], which is included in the overall effective area of the instrument.}
\label{fig:arf}
\end{figure}

\begin{itemize}
\item The first part will model the total effective area of the detector as a function of energy (Figure~\ref{fig:arf} -- \textit{Left}). In the case of the X-IFU, this includes the effective area of the mirror, the transmission of the thermal filters [\citen{Barbera2016Filters}] (Figure~\ref{fig:arf} -- \textit{Right}), the absorber stopping power of the pixel (also known as the quantum efficiency or QE) and the geometrical filling factor of the detector. This information is usually contained in the Ancillary Response Function (ARF) file.

\item The second part will model the energy redistribution of a given detector, which gives the probability of a photon of energy $E$ to fall into the instrumental channel of energy $E'$. The core part of the redistribution function is typically a Gaussian function centred on $E$, whose full-width half maximum (FWHM) is by definition the energy resolution of the detector at the given energy.  The energy redistribution of the pixel is contained in the Redistribution Matrix Function (RMF) file.
\end{itemize}

In this study, we consider the calibration of the instrumental responses for the X-IFU. The X-IFU is required to know the energy resolution to better than 0.15~eV up to 7~keV, the normalisation of the effective area with an absolute error lower than 4\% at 1~keV (TBC) and with relative error (i..e, knowledge of the shape of the effective area) below 3\% over the 0.2 -- 10~keV bandpass. 

Strategies for calibration of the instrumental response include ground calibration of the telescope and the filters, full-detector measurement of the quantum efficiency (i.e., each pixel is given the average quantum efficiency measured over the entire detector array TBC) and pixel-by-pixel measurements of the energy resolution. Yet, these approaches are not immune to inducing small systematic effects in the data. For instance, by assuming an average value of the QE over the entire detector, small pixel-to-pixel variations in the stopping power will in practice be overlooked. Likewise, although the energy resolution will be measured for each detector, this will only be done at several specific energies over the bandpass and interpolated, thus leaving residual uncertainties in the energy resolution of the pixels. Due to limited calibration time, the systematic errors introduced by these uncertainties may not be fully investigated through ground calibration. Simulations are thus required to quantify these effects on the science performed by the X-IFU.

\section{Numerical validation of calibration requirements}
\label{sec:num}

To investigate the effects of uncertainties in the instrumental response on the performances of the instrument, we have developed a dedicated pipeline using modified responses and based on an iterative approach. The first step of the simulation is to identify a representative science case which is most affected by an error in the physical parameter under investigation (e.g., the QE or the energy resolution). Starting from a realistic range of uncertainties on the subsystems, this can be estimated by computing a set of changes in the response functions and deducing where the highest effect on the science will be (e.g., on specific lines, on an energy band, on the continuum). One drawback of this approach is that more than one science objective may be driving the calibration, in which case, multiple studies may be needed. Once a relevant target is chosen, the simulation pipeline can be summarized as follows:

\begin{itemize}
\item The science case under consideration is represented by a typical input spectrum generated using XSPEC [\citen{Arnaud1996XSPEC}] and seen by a perfect instrument (unitary response). To study only the systematic effects related to the response functions, this spectrum is generated using unrealistically high exposure times. 
\item Using the realistic variations of the parameter under study, we generate a random value of the modified instrumental response function. The input spectrum is convolved with this response to obtain a realistic observation by the instrument.
\item The spectrum is fitted using C-statistics [\citen{Cash1979}] with the same input model but using instead the original value of the response function (i.e., not modified) to estimate the systematic error on the recovered parameters.
\end{itemize}

By repeating this process a large number of times, the overall effect of the instrumental response can be assessed. Physical parameters such as the QE may however depend on multiple other variables (e.g., thickness or composition of the absorbers, geometry...). To simplify the interpretation of the results, only changes in one of these variables are considered in this study.  Once the effects are quantified, they can be compared to the current science requirements for the X-IFU. If the errors are not compatible, mitigation techniques or improvements on subsystems may be required. This process can easily be generalised to other physical quantities and therefore provides an interesting tool to verify the coherence of the calibration requirements formulated for the X-IFU. 

\section{Quantum efficiency}
\label{sec:qe}

The Quantum Efficiency (QE) of a TES detector is defined as the overall stopping power of the pixel as a function of energy, and is \textit{de facto} related to the thickness of the absorbers placed above the TES. We investigate in this section the effect of changes in the absorber thickness on the instrumental response, at a single-pixel level. 

\subsection{Realistic variations in the detector's quantum efficiency}

We consider here that typical X-IFU detectors are composed of a 1.7~\textmu m Au and 4.2~\textmu m Bi absorber bi-layer with a geometrical filling factor of 96.8\% [\citen{Smith2016Pix}]. In the case of the gold layer, deposition is expected to be homogeneous within a pixel, with only small gradients in gold thickness across the wafer (estimated $\leq 5$\% TBC). Bismuth will on the contrary have inhomogeneous distribution within pixels [\citen{Sadleir2006Bi}], with differences up to 10\% (presence of `craters' and `valleys').  As long as the variation of bismuth is within this limit, the quantum efficiency at a given energy will remain within a linear range of variations over the bandpass of better than 1\%. As such, we assume here that the average pixel quantum efficiency should be (to the first order) comparable to the quantum efficiency of a pixel with an average value of bismuth thickness. In the rest of this study, changes in Bi are therefore represented by an `average' change in Bi thickness over the full pixel. This result remains to be confirmed by more accurate measurements on representative TESs. 

\begin{figure}
\centering
\includegraphics[width=0.495\textwidth]{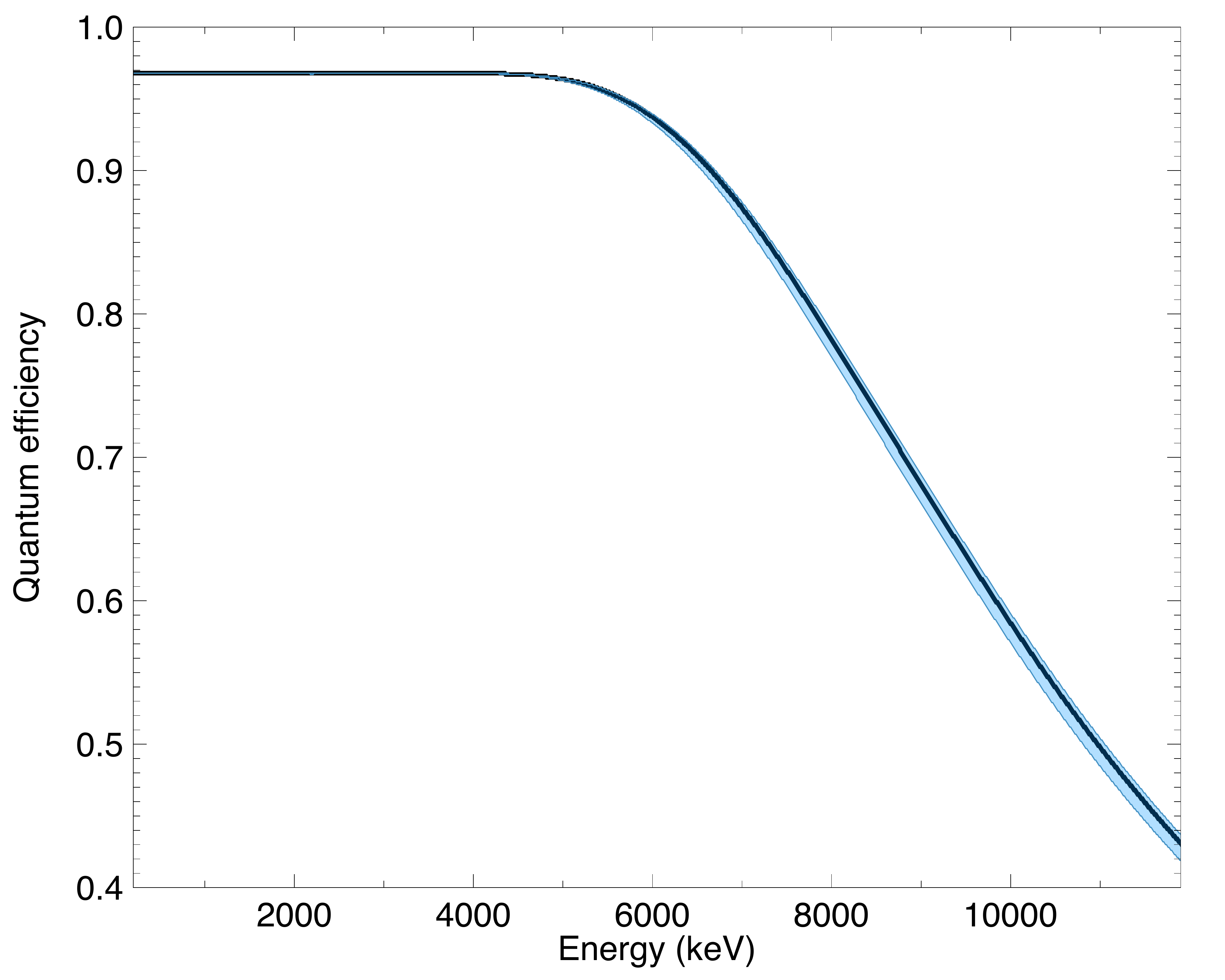}
\includegraphics[width=0.495\textwidth]{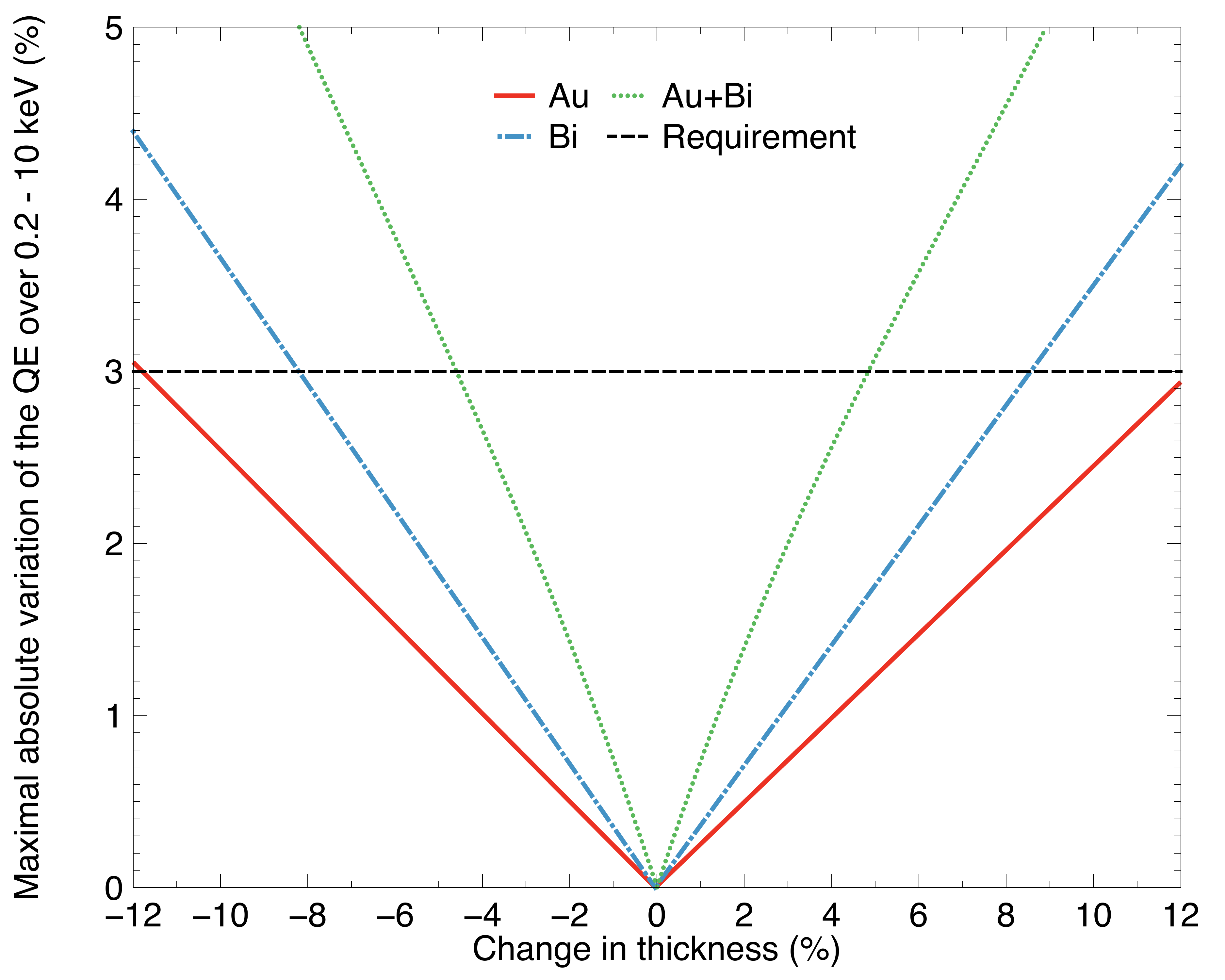}
\caption{(\textit{Left}) Quantum efficiency of the absorber with 1.7~\textmu m Au and 4.2~\textmu m Bi with a geometrical filling factor of 96.8\%. The blue envelope represents the change in QE for a $\pm 0.05$~\textmu m change in Au and in Bi, used as $\pm 1 \sigma$ envelope to generate realistic variations in the response. (\textit{Right}) Maximal absolute change in the absorber quantum efficiency over the 0.2 -- 10~keV bandpass as a function of the change (in \%) in the absorber thickness for Au (red solid line), Bi (blue dash-dotted line) and both in Au and in Bi (green dotted line). The horizontal dashed line recalls the 3\% requirement.}
\label{fig:abs}
\end{figure}

For both materials, changes in absorber thickness affect mainly the high-energy QE of the pixel, notably above 6~keV (see Figure~\ref{fig:abs} -- \textit{Left}). To remain below the required 3\% change over the 0.2 -- 10~keV bandpass at a pixel level, variations in the absorber thickness should not be higher than $\pm 0.20$~\textmu m in gold and $\pm 0.33$~\textmu m for bismuth i.e., respectively $\sim$~12\% and 8\% of their original value, as shown in Figure~\ref{fig:abs} (\textit{Right}). In fact, changes in either of these two materials have similar effects on the overall stopping power, as no atomic edge is present below 10~keV. Variations in the QE remain below the requirement as long as total relative variations of both absorbers are below $\sim$~4\% (Figure~\ref{fig:abs} -- \textit{Right}). Although bismuth is more likely to have higher absolute thickness variations, we assume that (in average over a pixel) these variations should be comparable to that of the gold gradients. In the rest of the study, we therefore use the same absolute change in thickness for both absorbers (i.e., $\Delta$Au=$\Delta$Bi in \textmu m). The same exercise has been conducted for comparison using the same relative variation for both absorbers (i.e., $\Delta$Au/Au=$\Delta$Bi/Bi in \%), with consistent results.

To investigate this effect on the performance of the X-IFU, we select as driving case the recovery of the physical properties of the hot plasma embedded in clusters of galaxies using the iron~K complex (5 -- 7~keV bandpass), which is likely to be affected by QE uncertainties above 6~keV. These measurements will be particularly critical to assess the dynamics of the cluster and its central object. In the current requirements of the instrument, iron abundance and temperature shall be recovered within 4\% of their original value, while the velocity broadening of the lines shall be known to better than 20~km/s. 

A typical emission spectrum of the thermal plasma was generated using XSPEC (\texttt{bapec} model [\citen{Smith2001APEC}]) for a local cluster ($z \sim 0.1$) with an iron abundance of $Z=0.3~Z_{\odot}$ (solar abundances as per [\citen{Anders1989Solar}]) and a turbulent velocity of 100~km/s. This spectrum is used as input of the numerical pipeline presented in Sect.~\ref{sec:num}. For each iteration of the pipeline, the modified response function is generated by randomly interpolating a value of the detector QE (Figure~\ref{fig:abs} -- \textit{Left}), assuming a normal distribution centred on the original function and with a standard deviation corresponding to the stopping power curves created for the same change of $\pm$~X~\textmu m in gold and bismuth thicknesses.

\begin{figure} [t]
\centering
\includegraphics[width=0.49\textwidth]{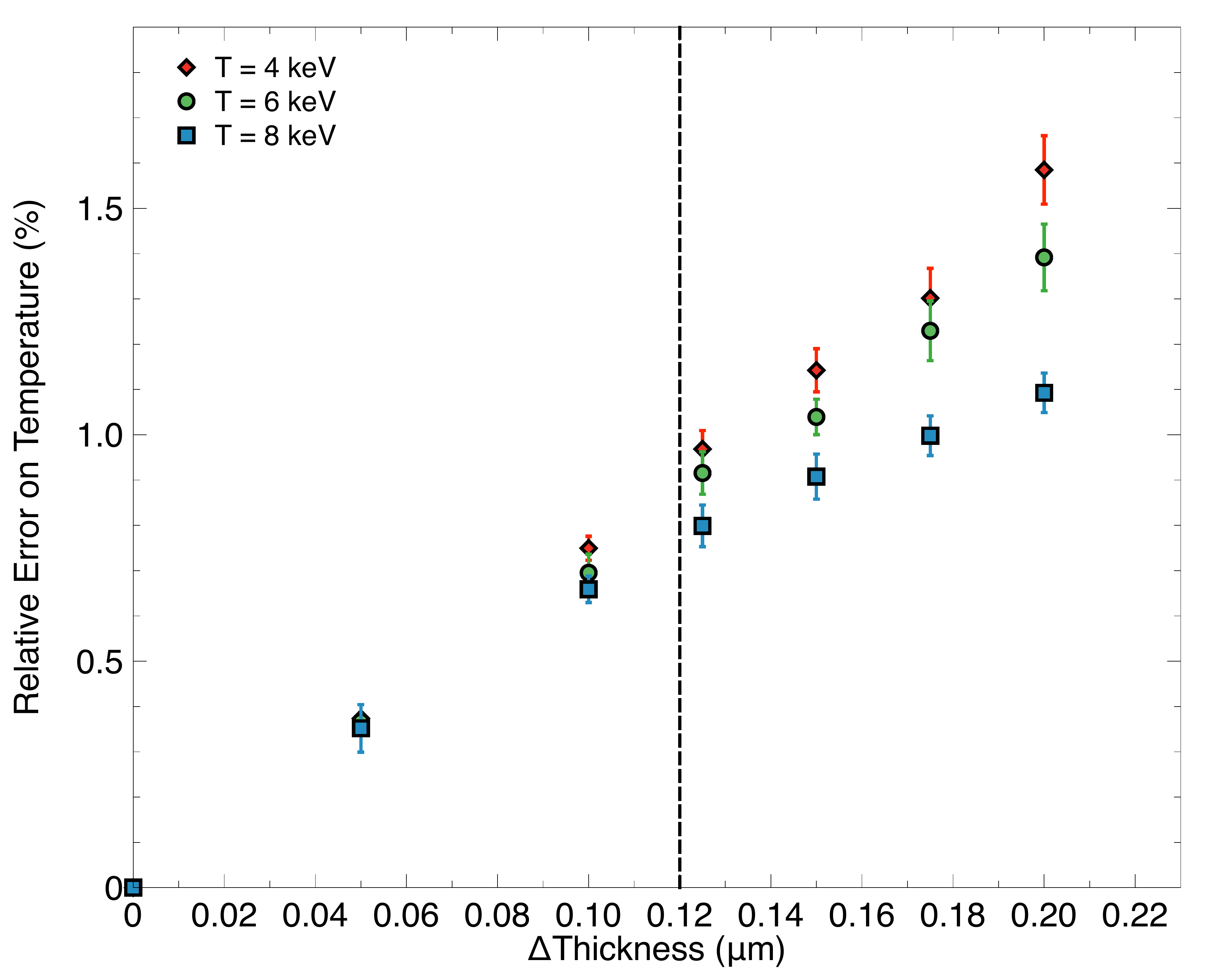}
\includegraphics[width=0.49\textwidth]{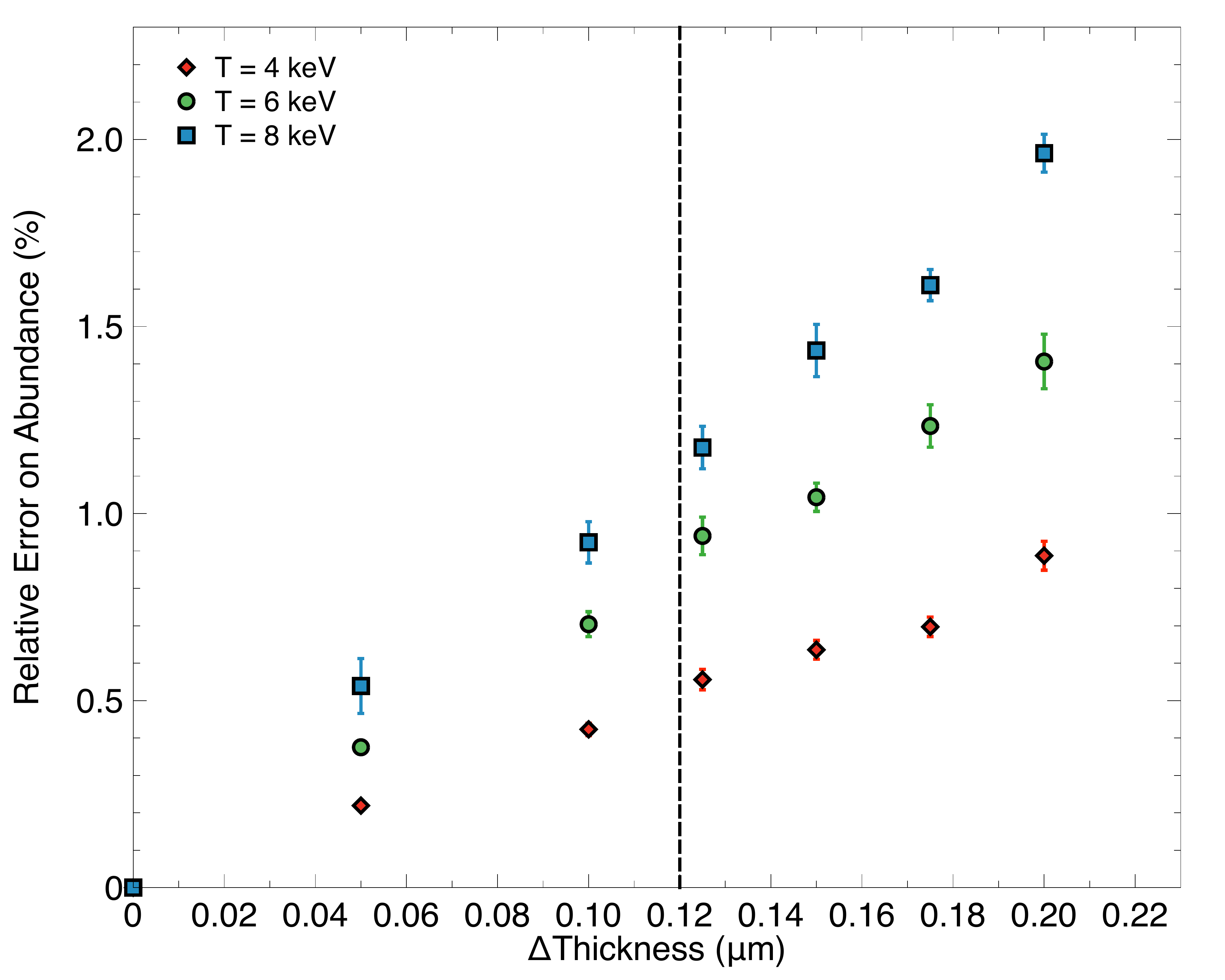}
\includegraphics[width=0.49\textwidth]{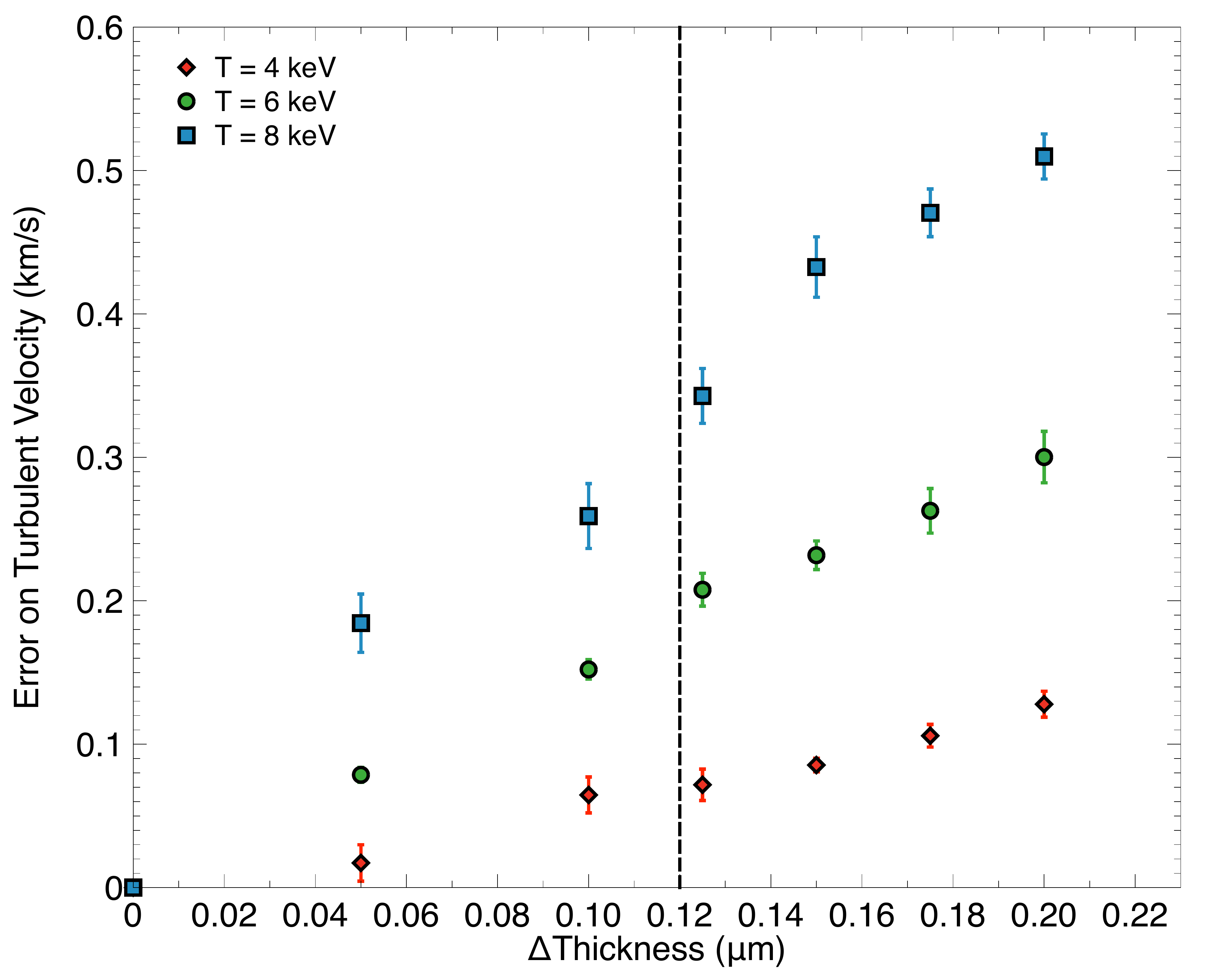}
\caption{Effect of a change in detector thickness (in \textmu m in gold and in bismuth) on the recovery of different physical parameters in terms of relative error (\%) with respect to their original value for temperature (\textit{Top left}) and iron abundance (\textit{Top right}) and in km/s for the turbulent velocity (\textit{Bottom}). Simulations are done for a cluster temperature of 4 (red diamonds), 6 (green circles) and 8~keV (blue squares) using 10\,000 iterations of the pipeline. The vertical dashed line indicates the change in Au and in Bi (\textmu m) such that the maximal change in the absolute QE reaches 3\% over 0.2 -- 10~keV.}
\label{fig:qe}
\end{figure}

\subsection{Impact on the observations}

The previous pipeline was applied to a set of `toy' model galaxy clusters, with hot gas temperatures of 4, 6 and 8~keV. Figure~\ref{fig:qe} shows the cluster's temperature, iron metallicity and turbulent speed over the 5~--~7~keV bandpass as function of the change in thickness used to modify the QE (both Au and Bi). In this case, the 3\% requirement is achieved for a change in 0.12~\textmu m in both layers, which represents a total $\sim$~4\% in the total thickness of the absorber. We find that abundances and temperature are most affected, with errors up to 1\% for variations of the order of $\pm 0.1$~\textmu m in both Au and Bi, while systematic errors on the velocity broadening are minimal ($\leq$1~km/s, see Figure~\ref{fig:qe} -- \textit{Bottom}). This can be explained by the shape of the changes in QE, which mostly affect the continuum of the spectrum rather than the broadening of the lines. In both cases, these effects are well below the current requirements. 

Given the current estimates expected on the thickness of the absorbers, the systematic effect on the observation of the iron K complex in galaxy clusters should be low per pixel. Most of the changes between pixels will besides be either slow gradients for gold, or averaged over the pixel for bismuth. Since data within neighbouring pixels will likely be grouped during post-processing to increase the signal-to-noise ratio, the effect shown Figure~\ref{fig:qe} will be further reduced within binned regions (for Bi notably). Even if single distant pixels are used (e.g., to compute structure functions to determine the power spectrum of the turbulence [\citen{ZuHone2016Turb}]), the expected variations across the wafer will be low: systematic effects should therefore be negligible with respect to statistical errors, which are of the order of a few percent for typical cluster observations. Experimental verification of the residual gradients across the wafer (for gold) and within pixels (for bismuth) should nevertheless be performed. As shown by the small sample of `toy' model clusters, results obtained here are strongly related to the choice of the input model, and ultimately to the choice of the science case. Further studies on other science objectives and other contributors to the effective area (e.g., filters) are therefore needed to validate the calibration requirements.

\section{Energy resolution}
\label{sec:res}

We investigate here the effect of uncertainties in the energy resolution in the instrumental response. 

\subsection{Impact on the energy resolution}

The energy resolution for the X-IFU will be measured at the pixel level. It will be calibrated by measuring the Line Spread Function (or LSF) of the detectors over the bandpass of the instrument (using e.g., crystal-cut mono-chromators or electron beam ion traps) and interpolated using polynomials [\citen{Eckart2018Gain}]. The Gaussian core LSF of the X-IFU micro-calorimeters should dominate the total LSF. Current response files (RMF) therefore use a Gaussian spectral redistribution function, and define its energy resolution as its FWHM $\Delta E_0$.  The knowledge of this energy resolution is required to be below 0.15~eV (up to 7~keV) for each pixel.  Small deviations of the energy resolution are however possible either due to the interpolation over the energy band or to statistical errors on the line fit used for ground calibration.

Uncertainties in the energy resolution will affect measurements of the line broadening in astrophysical sources. We select as driving science objective the measurement of the iron K complex velocity broadening (crucial to understand turbulence mechanisms in galaxy clusters), which requires an accuracy lower than 20~km/s.  The science case is modelled using the same \texttt{bapec} model presented in Sect.~\ref{sec:qe}, with a gas temperature of 4~keV. In the same way as for the QE, the input spectrum generated using XSPEC is used as input of the numerical pipeline Sect.~\ref{sec:num}. Changes in the response affect in this case only the RMF file, which is tweaked to have a modified spectral resolution $\Delta E$ between 5 and 7~keV. 

\subsection{Impact on turbulent speed measurements and analytical model}

\begin{figure} [t]
\centering
\includegraphics[width=0.8\textwidth]{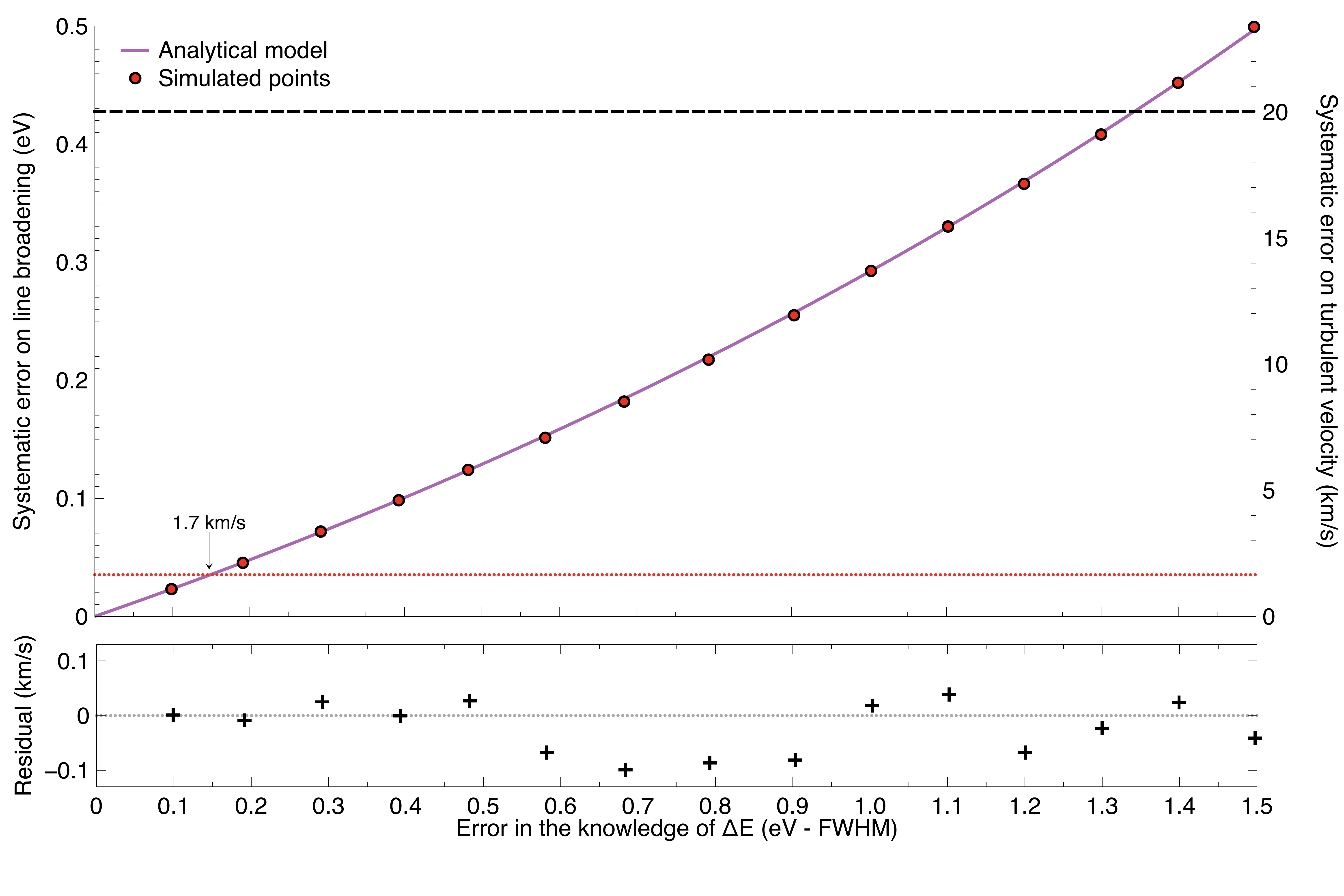}
\caption{Effect of an error in the knowledge of the energy resolution of a pixel (eV) on the measurement of the iron K line broadening (red dots), assuming a level of turbulence of $V_{\text{turb}}=100$~km/s. The analytical model of Equation~\ref{eq:vel} is over-plotted in purple. Degradation on the broadening is given both in eV and km/s (see Equation \ref{eq:conv}). The black dashed line gives the 20~km/s requirement, the red dotted line indicates the systematic error made for a 0.15~eV error in the knowledge of $\Delta E$. The bottom panel indicates the residuals (km/s) between the analytical model and the simulation.}
\label{fig:vel}
\end{figure}

The results provided by the simulations are shown in Figure~\ref{fig:vel}. For small variations of $\Delta E$, the impact of the energy resolution on the measurement is moderate. Even for uncertainties of the energy resolution of the order of 0.15~eV, the corresponding systematic error will be below 2~km/s. In an effort to generalise this study to other atomic lines, we tried to derive an analytical expression of the systematic degradation on the line broadening $\Delta \sigma_{\text{turb}}$ as a function of the error on $\Delta E$. To do so, assuming emission lines measured by the X-IFU are well represented by Gaussian, their total broadening $\sigma$ is given by:
\begin{equation}
\sigma^2 = \sigma^2_{\text{turb}} + \sigma^2_{\text{th}} + \sigma^2_{\text{res}}
\end{equation}
where $ \sigma^2_{\text{th}}$ is the thermal broadening of the line and $\sigma^2_{\text{res}} = \Delta E / \sqrt{8 \times ln(2)}$. The thermal broadening component is assumed to be fixed in this study, since temperature is frozen in the fit. Assuming an uncertainty  $\Delta \sigma_{\text{res}}$ on the energy resolution, for the same line, the error made on the turbulent velocity should be 
\begin{align}
|\Delta \sigma_{\text{turb}}| &= \sigma_{\text{turb}} - \sqrt{\sigma_{\text{turb}}^2 + \sigma_{\text{res,0}}^2 - \sigma_{\text{res}}^2}~~~\text{if}~\Delta \sigma_{\text{res}} \geq 0 \\
|\Delta \sigma_{\text{turb}}| &= - \sigma_{\text{turb}} + \sqrt{\sigma_{\text{turb}}^2 + \sigma_{\text{res,0}}^2 - \sigma_{\text{res}}^2}~~~\text{if}~\Delta \sigma_{\text{res}} \leq 0
\label{eq:vel}
\end{align}
where $\sigma_{\text{res,0}} =2.5$~eV and $\sigma_{\text{res}} = \sigma_{\text{res,0}} + \Delta \sigma_{\text{res}}$ As demonstrated in Figure~\ref{fig:vel}, the corresponding analytical model provides consistent results with the simulation. The systematic error on the broadening is converted in terms of km/s by using 
\begin{equation}
\Delta V_{\text{turb}} = \Delta \sigma_{\text{turb}} \frac{c}{E_w}
\label{eq:conv}
\end{equation}
where $c$ is the speed of light and $E_w$ the profile-weighted energy of the complex. Once again, this systematic effect will be higher when single pixels are considered, but should be averaged whenever multiple pixels are binned together. For purely Gaussian lines, Equation~\ref{eq:vel} should provide accurate first-order estimates of the systematic error on the turbulent velocity over the bandpass. Overall, these results suggest that the contribution of systematic errors in the calibration requirement for the energy resolution should be low. Statistics of the line will likely play a much larger part in the total error contribution for science cases involving line broadening measurements, notably in the case of weak lines.

\section{Conclusion}
In this paper we have presented a numerical approach to test and verify some of the calibration requirements for the X-ray Integral Field Unit. A specific look was given to the calibration of the instrumental response. Starting from realistic variations of the physical parameters (e.g., thickness of TES absorbers, LSF) and using modified response files, the pipeline was used to investigate the calibration requirement for the quantum efficiency and the energy resolution. Notably the effect of uncertainties in these parameters was quantified on the driving science case of iron K measurements in galaxy clusters. Overall, we have demonstrated that for the current design of the instrument, these systematics can be considered as second-order effects and will likely be averaged in binned pixel regions. However, most of the results obtained here are model-dependent. Further investigations by using multiple science cases along with actual measurements in the subsystems are now required to continue the validation of the X-IFU calibration requirements.

\bibliography{Calib_paper} 

\begin{thebibliography}{10}

\bibitem{Barret2016XIFU}
{Barret}, D., {Lam Trong}, T., {den Herder}, J.-W., {Piro}, L., {Barcons}, X.,
  {Huovelin}, J., {Kelley}, R., {Mas-Hesse}, J.~M., {Mitsuda}, K., {Paltani},
  S., {Rauw}, G., {Ro{\.Z}anska}, A., {Wilms}, J., {Barbera}, M., {Bozzo}, E.,
  {Ceballos}, M.~T., {Charles}, I., {Decourchelle}, A., {den Hartog}, R.,
  {Duval}, J.-M., {Fiore}, F., {Gatti}, F., {Goldwurm}, A., {Jackson}, B.,
  {Jonker}, P., {Kilbourne}, C., {Macculi}, C., {Mendez}, M., {Molendi}, S.,
  {Orleanski}, P., {Pajot}, F., {Pointecouteau}, E., {Porter}, F., {Pratt},
  G.~W., {Pr{\^e}le}, D., {Ravera}, L., {Renotte}, E., {Schaye}, J.,
  {Shinozaki}, K., {Valenziano}, L., {Vink}, J., {Webb}, N., {Yamasaki}, N.,
  {Delcelier-Douchin}, F., {Le Du}, M., {Mesnager}, J.-M., {Pradines}, A.,
  {Branduardi-Raymont}, G., {Dadina}, M., {Finoguenov}, A., {Fukazawa}, Y.,
  {Janiuk}, A., {Miller}, J., {Naz{\'e}}, Y., {Nicastro}, F., {Sciortino}, S.,
  {Torrejon}, J.~M., {Geoffray}, H., {Hernandez}, I., {Luno}, L., {Peille}, P.,
  {Andr{\'e}}, J., {Daniel}, C., {Etcheverry}, C., {Gloaguen}, E., {Hassin},
  J., {Hervet}, G., {Maussang}, I., {Moueza}, J., {Paillet}, A., {Vella}, B.,
  {Campos Garrido}, G., {Damery}, J.-C., {Panem}, C., {Panh}, J., {Bandler},
  S., {Biffi}, J.-M., {Boyce}, K., {Cl{\'e}net}, A., {DiPirro}, M., {Jamotton},
  P., {Lotti}, S., {Schwander}, D., {Smith}, S., {van Leeuwen}, B.-J., {van
  Weers}, H., {Brand}, T., {Cobo}, B., {Dauser}, T., {de Plaa}, J., and
  {Cucchetti}, E., ``{The Athena X-ray Integral Field Unit (X-IFU)},'' in [{\em
  Space Telescopes and Instrumentation 2016: Ultraviolet to Gamma
  Ray}{\nolinebreak\hspace{0.1em}]},  {\em Proc. SPIE} {\bf 9905},  99052F
  (2016).

\bibitem{Pajot2018XIFU}
{Pajot}, F., {Lam Trong}, T., {den Herder}, J.-W., {Piro}, L., and {Cappi}, M.,
  ``{The Athena X-ray Integral Field Unit (X-IFU)},'' {\em Journal Of Low
  Temperature Physics}  (Jan. 2018).

\bibitem{Nandra2013Athena}
{Nandra}, K., {Barret}, D., {Barcons}, X., {Fabian}, A., {den Herder}, J.-W.,
  {Piro}, L., {Watson}, M., {Adami}, C., {Aird}, J., {Afonso}, J.~M., and
  et~al., ``{The Hot and Energetic Universe: A White Paper presenting the
  science theme motivating the Athena+ mission},'' {\em ArXiv e-prints}  (June
  2013).

\bibitem{Smith2016Pix}
{Smith}, S.~J., {Adams}, J.~S., {Bandler}, S.~R., {Betancourt-Martinez}, G.~L.,
  {Chervenak}, J.~A., {Chiao}, M.~P., {Eckart}, M.~E., {Finkbeiner}, F.~M.,
  {Kelley}, R.~L., {Kilbourne}, C.~A., {Miniussi}, A.~R., {Porter}, F.~S.,
  {Sadleir}, J.~E., {Sakai}, K., {Wakeham}, N.~A., {Wassell}, E.~J., {Yoon},
  W., {Bennett}, D.~A., {Doriese}, W.~B., {Fowler}, J.~W., {Hilton}, G.~C.,
  {Morgan}, K.~M., {Pappas}, C.~G., {Reintsema}, C.~N., {Swetz}, D.~S.,
  {Ullom}, J.~N., {Irwin}, K.~D., {Akamatsu}, H., {Gottardi}, L., {den Hartog},
  R., {Jackson}, B.~D., {van der Kuur}, J., {Barret}, D., and {Peille}, P.,
  ``{Transition-edge sensor pixel parameter design of the microcalorimeter
  array for the x-ray integral field unit on Athena},'' in [{\em Space
  Telescopes and Instrumentation 2016: Ultraviolet to Gamma
  Ray}{\nolinebreak\hspace{0.1em}]},  {\em Proc. SPIE} {\bf 9905},  99052H
  (July 2016).

\bibitem{Barbera2016Filters}
{Barbera}, M., {Argan}, A., {Bozzo}, E., {Branduardi-Raymont}, G.,
  {Ciaravella}, A., {Collura}, A., {Cuttaia}, F., {Gatti}, F., {Jimenez
  Escobar}, A., {Lo Cicero}, U., {Lotti}, S., {Macculi}, C., {Mineo}, T.,
  {Nuzzo}, F., {Paltani}, S., {Parodi}, G., {Piro}, L., {Rauw}, G.,
  {Sciortino}, L., {Sciortino}, S., and {Villa}, F., ``{Thermal Filters for the
  ATHENA X-IFU: Ongoing Activities Toward the Conceptual Design},'' {\em
  Journal of Low Temperature Physics}~{\bf 184},  706--711 (Aug. 2016).

\bibitem{Arnaud1996XSPEC}
{Arnaud}, K.~A., ``{XSPEC: The First Ten Years},'' in [{\em Astronomical Data
  Analysis Software and Systems V}{\nolinebreak\hspace{0.1em}]},  {Jacoby},
  G.~H. and {Barnes}, J., eds., {\em Astronomical Society of the Pacific
  Conference Series} {\bf 101},  17 (1996).

\bibitem{Cash1979}
{Cash}, W., ``{Parameter estimation in astronomy through application of the
  likelihood ratio},'' {\em Astrophysical Journal}~{\bf 228},  939--947 (Mar.
  1979).

\bibitem{Sadleir2006Bi}
{Sadleir}, J.~E., {Bandler}, S.~R., {Brekosky}, R.~P., {Chervenak}, J.,
  {Figueroa-Feliciano}, E., {Finkbeiner}, F., {Iyomoto}, N., {Kelley}, R.~L.,
  {Kilbourne}, C.~A., {King}, J.~M., {Porter}, F.~S., {Robinson}, I.~K.,
  {Saab}, T., and {Talley}, D.~J., ``{Bismuth X-ray absorber studies for TES
  microcalorimeters},'' {\em Nuclear Instruments and Methods in Physics
  Research A}~{\bf 559},  447--449 (Apr. 2006).

\bibitem{Smith2001APEC}
{Smith}, R.~K., {Brickhouse}, N.~S., {Liedahl}, D.~A., and {Raymond}, J.~C.,
  ``{Collisional Plasma Models with APEC/APED: Emission-Line Diagnostics of
  Hydrogen-like and Helium-like Ions},'' {\em Astrophysical Journal}~{\bf 556},
   L91--L95 (Aug. 2001).

\bibitem{Anders1989Solar}
{Anders}, E. and {Grevesse}, N., ``{Abundances of the elements - Meteoritic and
  solar},'' {\em Geochimica et Cosmochimica Acta}~{\bf 53},  197--214 (Jan.
  1989).

\bibitem{ZuHone2016Turb}
{ZuHone}, J.~A., {Markevitch}, M., and {Zhuravleva}, I., ``{Mapping the Gas
  Turbulence in the Coma Cluster: Predictions for Astro-H},'' {\em
  Astrophysical Journal}~{\bf 817},  110 (Feb. 2016).

\bibitem{Eckart2018Gain}
{Eckart}, M.~E., {Adams}, J.~S., {Boyce}, K.~R., {Brown}, G.~V., {Chiao},
  M.~P., {Fujimoto}, R., {Haas}, D., {den Herder}, J.~W., {Ishisaki}, Y.,
  {Kelley}, R.~L., {Kilbourne}, C.~A., {Leutenegger}, M.~A., {McCammon}, D.,
  {Mitsuda}, K., {Porter}, F.~S., {Sato}, K., {Sawada}, M., {Seta}, H.,
  {Sneiderman}, G.~A., {Szymkowiak}, A.~E., {Takei}, Y., {Tashiro}, M.,
  {Tsujimoto}, M., {de Vries}, C.~P., {Watanabe}, T., {Yamada}, S., and
  {Yamasaki}, N.~Y., ``Ground calibration of the astro-h (hitomi) soft x-ray
  spectrometer,'' {\em Journal of Astronomical Telescopes, Instruments, and
  Systems}~{\bf 4},  4 -- 4 -- 22 (2018).

\end{thebibliography}
\bibliographystyle{spiebib} 

\end{document}